\documentclass[prd,aps,12pt,nofootinbib,superscriptaddress]{revtex4-1}
\usepackage{amsmath, amssymb, amsfonts, amsthm}
\usepackage[usenames,dvipsnames,svgnames,table]{xcolor}
\usepackage{graphicx}
\usepackage{subfigure}

\makeatletter
\newcommand{\Rmnum}[1]{\expandafter\@slowromancap\romannumeral #1@}
\makeatother

\newcommand{\beq}{\begin{equation}}
\newcommand{\eeq}{\end{equation}}

\newcommand{\ber}{\begin{eqnarray}}
\newcommand{\eer}{\end{eqnarray}}

\def\nn{\nonumber}
\def\d{{\rm d}}

\def\pa{\partial}

\def\l{\Lambda}
\def\etal{et. al.}
\def\f{\frac}
\def\l{\left}
\def\r{\right}
\def\ie{{\it i.e.~}}

\def\mnras{{Mon.\@ Not.\@ Roy.\@ Ast.\@ Soc.\ }}

\def\prl{{Phys.\@ Rev.\@ Lett.\ }}
\def\prd{{Phys.\@ Rev.\@ D\ }}

\def\plb {{Phys.\@ Lett.\@ B\ }}

\begin{document}

\newcolumntype{L}[1]{>{\raggedright\arraybackslash}p{#1}}
\newcolumntype{C}[1]{>{\centering\arraybackslash}p{#1}}
\newcolumntype{R}[1]{>{\raggedleft\arraybackslash}p{#1}}

\title{A new recipe for $\Lambda$CDM}

\author{Varun Sahni}\email{varun@iucaa.in}
\affiliation{Inter-University Centre for Astronomy and Astrophysics, Pune, India}
\author{Anjan A Sen}\email{aasen@jmi.ac.in}
\affiliation{Centre for Theoretical Physics, Jamia Millia Islamia, New Delhi-110025, India}

\bigskip

\begin{abstract}
It is well known that a canonical scalar field is able to describe either dark matter
or dark energy but not both. We demonstrate that a non-canonical scalar field can describe
{\em both} dark matter and dark energy within a unified setting. We consider the simplest extension
of the canonical Lagrangian ${\cal L} \propto X^\alpha - V(\phi)$ where
$\alpha \geq 1$ and $V$ is a sufficiently flat potential. 
In this case 
the kinetic term in
the Lagrangian behaves just like a perfect fluid, 
whereas the potential term mimicks the cosmological constant. 
For very
 large values, $\alpha \gg 1$, the equation of state of the kinetic term 
drops to zero 
and the universe expands as $\Lambda$CDM.
The velocity of sound in this model, and the associated 
gravitational clustering, is sensitive to the value of $\alpha$.
For very large values of $\alpha$ the clustering properties of our model
resemble those of
 cold dark matter (CDM). But for smaller values of $\alpha$,
gravitational clustering on
small scales is 
suppressed,
and our model has properties resembling those of warm dark matter (WDM).
Therefore our non-canonical model
 has an interesting new property: while the background universe
expands like $\Lambda$CDM,
its clustering properties can resemble those of either cold or warm dark matter.
\end{abstract}

\maketitle


\section{Introduction}\label{sec: intro}
Ever since the discovery that high redshift type Ia supernovae supported
 an accelerating
universe, concordance cosmology or $\Lambda$CDM, has come to dominate popular thinking.
Although issues relating to the smallness of $\Lambda$ have given rise to several 
rival models of cosmic acceleration \cite{review,ss06} there is no doubt that, despite some recent evidence to
the contrary \cite{bao_2014,sss14,sarkar15,vms15}, $\Lambda$CDM agrees 
well with a large set of cosmological observations \cite{planck_2015}.

As its name suggests, $\Lambda$CDM consists of two components: the cosmological constant, $\Lambda$,
and {\em cold dark matter} (CDM). Despite its enormous success in explaining observations,
the origin of $\Lambda$ is not known. It may simply be a residual vacuum fluctuation, although
quantum field theory usually predicts much larger values, and the cosmological constant problem
remains unresolved.
As concerns dark matter, mainstream thinking usually assumes it to be a non-baryonic
relic of the big bang
\cite{kt,dark_matter} but other explanations can also be found in the literature
\cite{sahni,mond,peebles99,sahni-wang,hu00,matos00,zeldovich}. Furthermore,
since $96\%$ of the content of the universe is of unknown
origin, attempts have been made to describe {\em both dark matter and dark energy} within
a unified setting.
The Chaplygin gas (and its subsequent generalization) belongs to this category of models, since its equation of state (EOS)
behaves like pressureless dust at early times and like a $\Lambda$-term at late times \cite{CG};
 also see \cite{tirth}.
Unfortunately the Chaplygin gas has problems with gravitational clustering and so falls
short of describing the real universe \cite{CG_pert,sandvik04,bertacca08,bertacca11,asen14}.

In this paper we show that a unified description of dark matter and dark energy
can emerge from non-canonical scalar fields; see \cite{scherrer04,tejedor_feinstein,bertacca07} for earlier work in
this direction. 
These fields possess an additional degree of freedom (encoded in the parameter $\alpha$)
which allows a scalar field rolling along a flat potential to behave like
a two component fluid consisting of an almost pressureless kinetic component (dark matter)
and a cosmological constant.
For large values of $\alpha$
the equation of state of the kinetic component drops to zero and 
the universe expands
like $\Lambda$CDM. 
Non-canonical scalars cluster on small scales, thereby
providing us with a realistic model of an accelerating universe consisting of dark matter
and dark energy.
For very large values of $\alpha$ the kinetic component clusters like cold dark matter,
whereas for smaller $\alpha$ values, clustering in our model resembles warm dark matter.

\section{Non-canonical scalars and $\Lambda$CDM}\label{sec: scalar}

Perhaps the simplest generalisation of the canonical scalar field Lagrangian density
\beq
{\cal L}(X,\phi) = X -\; V(\phi), ~~~~X = \frac{1}{2}{\dot\phi}^2
\label{eqn: Lagrangian0}
\eeq
which preserves the second order nature of the field equations, is 
the non-canonical 
Lagrangian \cite{Mukhanov-2006,tejedor_feinstein,sanil-2008,sanil13,wands12}
\beq
{\cal L}(X,\phi) = X\l(\frac{X}{M^{4}}\r)^{\alpha-1} -\; V(\phi), 
\label{eqn: Lagrangian}
\eeq
where $M$ has dimensions
of mass while $\alpha$ is dimensionless. When $\alpha = 1$
the k-essence Lagrangian (\ref{eqn: Lagrangian}) reduces to 
(\ref{eqn: Lagrangian0}).

We shall be working in the spatially flat Friedmann-Robertson-Walker (FRW) universe
\beq
\d s^2 = \d t^2-a^{2}(t)\; \l[\d x^2 + \d y^2 + \d z^2\r],
\label{eqn: FRW}
\eeq
for which the energy-momentum tensor has the form
\beq
T^{\mu}_{\;\:\;\nu} = \mathrm{diag}\l(\rho_{_{\phi}}, -p_{_{\phi}}, - p_{_{\phi}}, - p_{_{\phi}}\r),
\eeq
where the energy density, $\rho_{_{\phi}}$, and pressure, $p_{_{\phi}}$, are given by
\ber
\rho_{_{\phi}} &=& \l(\f{\pa {\cal L}}{\pa X}\r)\, (2\, X)- {\cal L}\label{eqn: rho-phi},\\
p_{_{\phi}} &=& {\cal L}\label{eqn: p-phi}.
\eer
Substituting for ${\cal L}$ from (\ref{eqn: Lagrangian}) into (\ref{eqn: rho-phi}) and (\ref{eqn: p-phi})
one gets
\ber
\rho_{_{\phi}} &=& \l(2\alpha-1\r)X\l(\frac{X}{M^{4}}\r)^{\alpha-1} +\;  V(\phi),\nonumber\\
p_{_{\phi}} &=& X\l(\frac{X}{M^{4}}\r)^{\alpha-1} -\; V(\phi), ~~
\label{eqn: p-model}
\eer
which reduces to the canonical form
$\rho_{_{\phi}} = X + V$, ~$p_{_{\phi}} = X - V$ when $\alpha = 1$.
The two Friedmann equations are
\ber
H^{2} &=& \frac{8 \pi G}{3}\l[\l(2\alpha-1\r)X\l(\frac{X}{M^{4}}\r)^{\alpha-1} +\;
V(\phi)\r]~,\label{eqn: FR-eqn1 model}\\
\frac{\ddot{a}}{a} &=& -\frac{8 \pi G}{3}\l[\l(\alpha + 1\r)X\l(\frac{X}{M^{4}}\r)^{\alpha-1} -\;  V(\phi)\r]~,
\label{eqn: FR-eqn2 model}
\eer

where $\phi(t)$ satisfies the equation of motion
\beq
{\ddot \phi}+ \f{3\, H{\dot \phi}}{2\alpha -1} + \l(\f{V'(\phi)}{\alpha(2\alpha -1)}\r)\l(\f{2\,M^{4}}{{\dot \phi}^{2}}\r)^{\alpha - 1} =\; 0,
\label{eqn: EOM-model}
\eeq
which reduces to the standard canonical form
${\ddot \phi}+ 3\, H {\dot \phi} + V'(\phi) = 0$ when $\alpha =1$.

Consider the equation of motion (\ref{eqn: EOM-model}) for the simplest case when
$V(\phi) = \Lambda/8\pi G$. Setting $V'=0$ in (\ref{eqn: EOM-model}) one finds 
\beq
{\ddot \phi} = -\f{3\, H{\dot \phi}}{2\alpha -1}~,
\eeq
which is easily integrated to give
\beq
{\dot\phi} \propto a^{-\frac{3}{2\alpha - 1}}~, 
\label{eq:phidot}
\eeq
and which reduces to the canonical result, ${\dot\phi} \propto a^{-3}$, when $\alpha = 1$.
Substituting for $X \equiv {\dot\phi}^2/2$ from (\ref{eq:phidot}) into (\ref{eqn: p-model}) one 
readily finds (we have set $8\pi G = 1$ for simplicity)
\ber
\rho_{_{\phi}} &=& \rho_X + \Lambda\label{eq:mixture1}\\
p_{_{\phi}} &=& p_X - \Lambda~,
\label{eq:mixture2}
\eer
with $\rho_X = \l(2\alpha-1\r)X\l(\frac{X}{M^{4}}\r)^{\alpha-1} 
 \equiv \rho_{0X} a^{-3(1+w)}$
and $p_X = w \rho_X$,
where
\beq
w = \frac{p_X}{\rho_X} = \frac{1}{2\alpha-1}~,
\label{eq:state}
\eeq
is the equation of state (EOS) of the kinetic component of the scalar field.
From (\ref{eq:mixture1}) \& (\ref{eq:mixture2}) it follows that the non-canonical scalar field
behaves like a mixture of two non-interacting perfect fluids: $\rho_X$ and $\Lambda$, where the
equation of state of $\rho_X$ is given by (\ref{eq:state}).
Substituting (\ref{eq:mixture1}) into (\ref{eqn: FR-eqn1 model}) one finds
\beq
H(z) = H_0\left\lbrack \Omega_{0X} (1+z)^{3(1+w)} + \Omega_\Lambda\right\rbrack^{1/2}
\label{eq:hubble}
\eeq
where $\Omega_{0X} = \frac{8\pi G\rho_{0X}}{3H_0^2}$ and $w$ is described by (\ref{eq:state}).

From (\ref{eq:state}) and (\ref{eq:hubble}) we find that the expansion history is very sensitive to the value of $\alpha$.
When $\alpha = 1$ the scalar field behaves like a mixture of `$\Lambda$ + stiff matter'.
For $\alpha = 2$ the expansion history mimicks `$\Lambda$ + radiation'.
For $\alpha \gg 1$, $w \to 0$ and $\rho_X \propto a^{-3}$,
consequently (\ref{eq:hubble}) describes $\Lambda$CDM
in this limit:
\beq
H(z) \simeq H_0\lbrack \Omega_{0m} (1+z)^3 + \Omega_\Lambda\rbrack^{1/2}~, ~~
{\rm where}~~ \Omega_{0m} \equiv \Omega_{0X}~.
\label{eq:hubble1}
\eeq

\begin{figure*}[!htb]
\begin{center}
\resizebox{220pt}{180pt}{\includegraphics{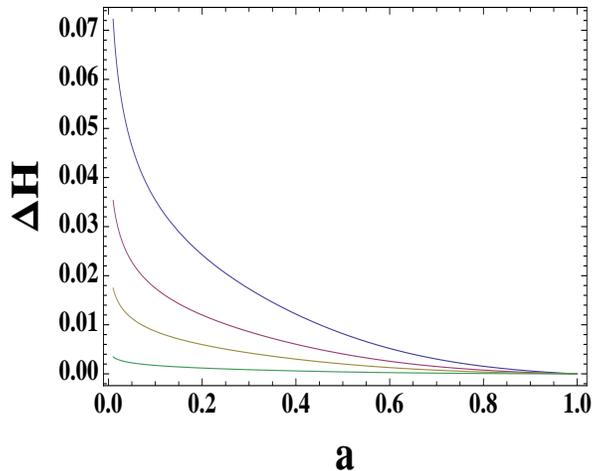}}
\end{center}
\caption{$\Delta H$ is plotted against the expansion factor, $a$,
for $\alpha = 50,100,200,1000$ (top to bottom). Here $\Delta H = (H-H_{\Lambda CDM})/H_{\Lambda CDM}$
and $a=1$ corresponds to the present epoch.}
\end{figure*}

The equation of state of the scalar field, $w_\phi = p_\phi/\rho_\phi$, is given by
\beq
1+ w_\phi(z) = \left (1+w\right )
\left\lbrack 1+\frac{\Omega_\Lambda}{\Omega_{0X}(1+z)^{3(1+w)}}\right\rbrack^{-1}
\label{eq:state1}
\eeq
where $w$ is described by (\ref{eq:state}).
We find that $w_\phi \simeq w$ when $z \gg 1$, while its current
value is
$w_{\phi,0} = (1+w)\left\lbrack 1+\frac{\Omega_\Lambda}{\Omega_{0X}}\right\rbrack^{-1}-1$.
From (\ref{eq:state}) and (\ref{eq:state1})
 one finds that for $\alpha \gg 1$, $w_\phi \simeq 0$ at $z \gg 1$,
and $w_{\phi,0} \simeq -\Omega_\Lambda$ at $z = 0$.
Thus for large values of $\alpha$, the EOS of the scalar field smoothly
interpolates between dust-like behaviour at high redshifts and a
negative value at present.\footnote{In the limit when $\alpha \to \infty$ our model
has properties resembling those of \cite{tejedor_feinstein}.}
In figure 1 we plot the fractional difference between the Hubble parameter $H(z)$ in our model from
$\Lambda$CDM for different values of the parameter $\alpha$ (but identical values of the matter density).
One can see for $\alpha \geq 10^3$ the deviation is less than $1\%$.
 Hence for such large values of $\alpha$ our model will be virtually
 indistinguishable from $\Lambda$CDM model
by observables measuring background cosmology alone.

We have assumed thus far that $V(\phi)$ is a constant. This however need not necessarily be the case. It is important to note that $\Lambda$CDM-like expansion
can also arise in the case of other potentials which are
flat. The reason for this is simple. Our treatment above was based on the
assumption that the last term in (\ref{eqn: EOM-model})
was negligibly small compared to the remaining two terms,
allowing the former to be neglected. This feature is shared by several flat potentials
some of which are described below.

\begin{itemize}

\item $V(\phi) = \frac{1}{2}m^2\phi^2$. In order to
 study this power-law potential, we first form an autonomous system of equations for our model. For Lagrangians of the form ${\cal L} = F(X) - V(\phi)$, De-Santiago et al \cite{santiago} have already constructed an autonomous system of equations involving the dimensionless variables
\begin{eqnarray}
x &=& \frac{\sqrt{2XF_{X} -F}}{\sqrt{3}m_{pl}H},\nonumber\\
y &=& \frac{\sqrt{V}}{\sqrt{3}m_{pl}H},\nonumber\\
w_{k} &=& \frac{F}{2XF_{X}-F},\nonumber\\
\sigma &=& -\frac{m_{pl}}{\sqrt{3|\rho_{k}|}}\frac{d \log V}{dt}.
\end{eqnarray}
Applying these variables to our Lagrangian given by (2), we get
\begin{eqnarray}
w_{k} &=& \frac{1}{\alpha -1},\nonumber\\
\Omega_{\phi} &=& x^2 + y^2 = 1,\nonumber\\
\gamma &=& 1 + w_{\phi} = x^{2}(1+w_{k}).
\end{eqnarray}
For $\alpha >>1$, $w_{k} \sim 0$ and one can safely approximate $\gamma \sim x^2$. Using the formulation prescribed in De-Santiago et al \cite{santiago}, we can now form an autonomous system of equations for $\gamma$ and $\sigma$:
\begin{eqnarray}
\gamma^{'} &=& 3\sigma (1-\gamma)\sqrt{\gamma} - 3\gamma(1-\gamma),\nonumber\\
\sigma^{'} &=& - 3\sigma^{2} \sqrt{\gamma}(\Gamma -1) + \frac{3}{2}\sigma \left(1-\sigma(1-\gamma)/\sqrt{\gamma}\right).
\end{eqnarray}
Here `{\it prime}' denotes derivative w.r.t $\log a$ and $\Gamma = \frac{V V''(\phi)}{V'(\phi)^2}$, so that $\Gamma = \frac{1}{2}$ for $V = \frac{1}{2}m^2\phi^2$. In order to
 solve this autonomous system, one requires initial conditions for $\gamma$ and $\sigma$. We set these at decoupling, $a \sim 10^{-3}$, assuming that initially the scalar field kinetic energy dominates over its potential energy, so that $w_{\phi} \sim \frac{1}{2\alpha -1}$. With $\alpha >>1$, we have $\gamma_{i} \sim 1$. Similarly one finds
 $\sigma_{i} \sim 0$ at decoupling. 
(One should note that $w_{\phi}$ is not exactly equal to
 zero initially, due to the large but 
finite value of $\alpha$.) With these initial conditions, the autonomous system of 
equations in (20) is evolved from decoupling until today, and the resultant
 behaviour of the equation of state for the scalar field, $w_{\phi}$,
 is shown in figure \ref{fig:2}. We find that the behaviour of $w_{\phi}$ in our model
 is very similar to that described by (18) for $\Lambda$CDM.

\begin{figure*}[!htb]
\begin{center} 
\resizebox{220pt}{180pt}{\includegraphics{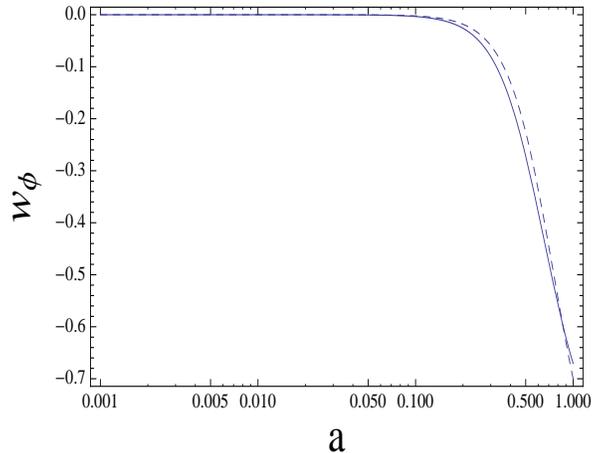}} 
\end{center}
\caption{The equation of state for the scalar field $w_{\phi}$ as a function of scale factor for $V(\phi) \sim \phi^2$ (Solid line). The dashed line is for equation of state described by (18) for $\Omega_{\Lambda} = 0.7$ and $\Omega_{0X} = 0.3$.} 
\label{fig:2}
\end{figure*}
 
\item Finally, an interesting example of a piece-wise
flat potential is the step potential
\beq
V(\phi) = A + B\tanh{\beta \phi}
\label{eq:step}
\eeq
where $A+B = V_{\rm initial}$
and $A-B = V_{\rm final}$. For
 $V_{\rm initial} \simeq 10^{64}{\rm GeV}^4$, $V_{\rm final} \simeq 10^{-47} {\rm GeV}^4$ 
this potential would interpolate between inflation at early times, and dark energy at late times,
and therefore might describe a model of Quintessenital-Inflation. We shall examine this
possibility in greater detail in a companion paper.

\end{itemize}

We therefore find that a single non-canonical scalar field can play the dual role of
dark matter and dark energy viz-a-viz the expansion history of the universe.
However in order to deepen the parallel between scalar field dynamics and $\Lambda$CDM
we also need to demonstrate that the field $\phi$ can cluster.
In order to do this we first note that
linearized scalar perturbations in a spatially flat FRW universe are described
by the line element \cite{Bardeen-1980,Kodama-1984,Mukhanov-1992}
\ber
\d s^2
&=& (1+2\, A)\,\d t ^2 - 2\, a(t)\, (\pa_{i} B )\; \d t\; \d x^i\,\nonumber\\
&-& a^{2}(t)\; \l[(1-2\, \psi)\; \delta _{ij}+ 2\, \l(\pa_{i}\, \pa_{j}E \r) \r]\,
\d x^i\, \d x^j\nn~.
\eer

The linearized Einstein's equation $\delta G^{\mu}_{\;\nu} = \kappa\, \delta T^{\mu}_{\;\nu}$ 
together with the perturbation equation for the scalar field give
\beq
\mathcal{R}_{_k}'' + 2\l(\frac{z'}{z}\r)\mathcal{R}_{_k}'  + c_{_s}^{2}k^{2}\mathcal{R}_{_k} = 0~,
\label{eqn: curvature pert eqn}
\eeq
where
\beq
z \equiv \frac{a\,\l(\rho_{_{\phi}}+ p_{_{\phi}}\r)^{1/2}}{c_{_s}H},
\eeq
$\mathcal{R}$ is the curvature perturbation 
\beq
\mathcal{R} \equiv \psi + \l(\frac{H}{\dot{\phi}}\r)\delta \phi~,
\eeq
and $\psi$, $\delta \phi$ correspond to the metric perturbation 
and the scalar field perturbation, respectively.
The derivative in (\ref{eqn: curvature pert eqn}) is taken with respect to the conformal time, 
$\eta = \int dt/a(t)$ and 
$c_s$ is the effective sound speed of perturbations
in the scalar field~\cite{Garriga-1999}
\beq
c_s^{2} \equiv \l[\f{\l({\pa {\cal L}}/{\pa X}\r)}{\l({\pa {\cal L}}/{\pa X}\r)
+ \l(2\, X\r)\, \l({\pa^{2} {\cal L}}/{\pa X^{2}}\r)}\r] ~.
\label{eq:sound_speed_def}
\eeq
Rewriting (\ref{eqn: curvature pert eqn})
in terms of the Mukhanov-Sasaki variable $u_{_{k}} \equiv z\,\mathcal{R}_{_k}$,
one gets
\beq
u_{_k}'' +  \l(c_{_s}^{2}k^{2} -   \frac{z''}{z}\r)u_{_k} = 0 ~.
\eeq
The key to our understanding of gravitational clustering is provided by
the sound speed. Substituting (\ref{eqn: Lagrangian}) into (\ref{eq:sound_speed_def})
we get
\beq
c_s^{2} = \f{1}{2\,\alpha - 1}.
\label{eqn: sound speed model}
\eeq
We therefore find that the sound speed is a constant, and that for $\alpha \gg 1$,
$c_{_S}^{2} \to 0$.
In other words, when the value of the non-canonical parameter $\alpha$ is large,
the sound speed vanishes, and the scalar field begins to behave like a pressureless
fluid.

An important property of our model follows from 
(\ref{eq:hubble1}) and (\ref{eqn: sound speed model}), namely, 
when $\alpha \gg 1$, the background universe
expands like $\Lambda$CDM, while
its clustering properties could resemble those of cold dark matter 
or even warm dark matter.
The non-canonical scalar therefore provides a unified prescription for 
dark matter and dark energy since both components are sourced by the same non-canonical scalar field.
We elaborate on this issue below.

\begin{figure*}[!htb]
\begin{center} 
\resizebox{200pt}{160pt}{\includegraphics{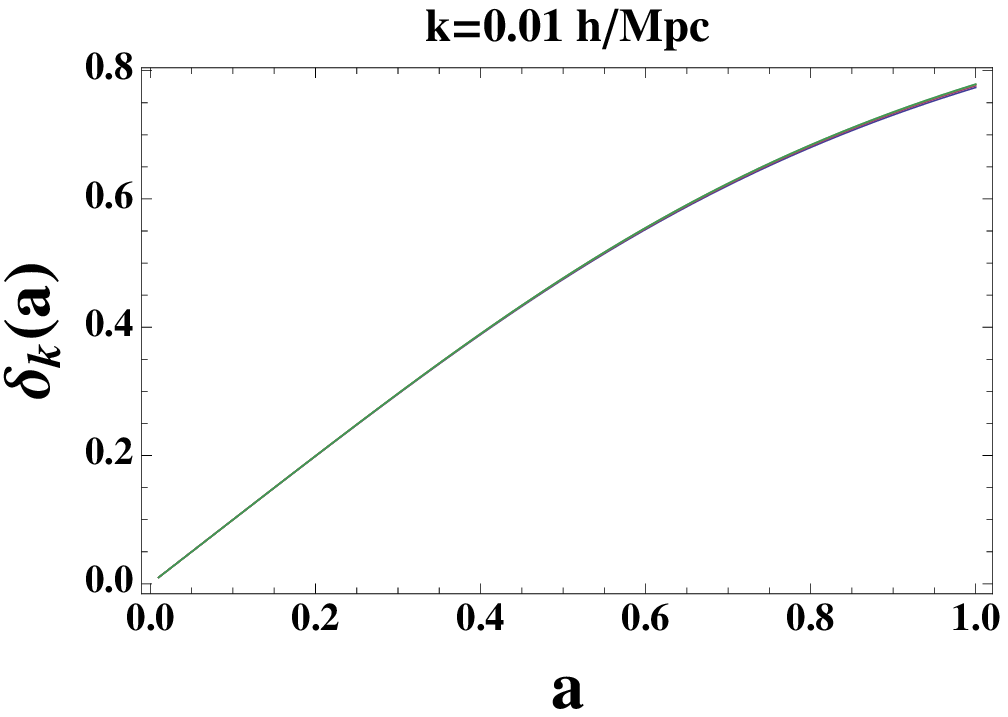}}
\hspace{1mm} \resizebox{200pt}{160pt}{\includegraphics{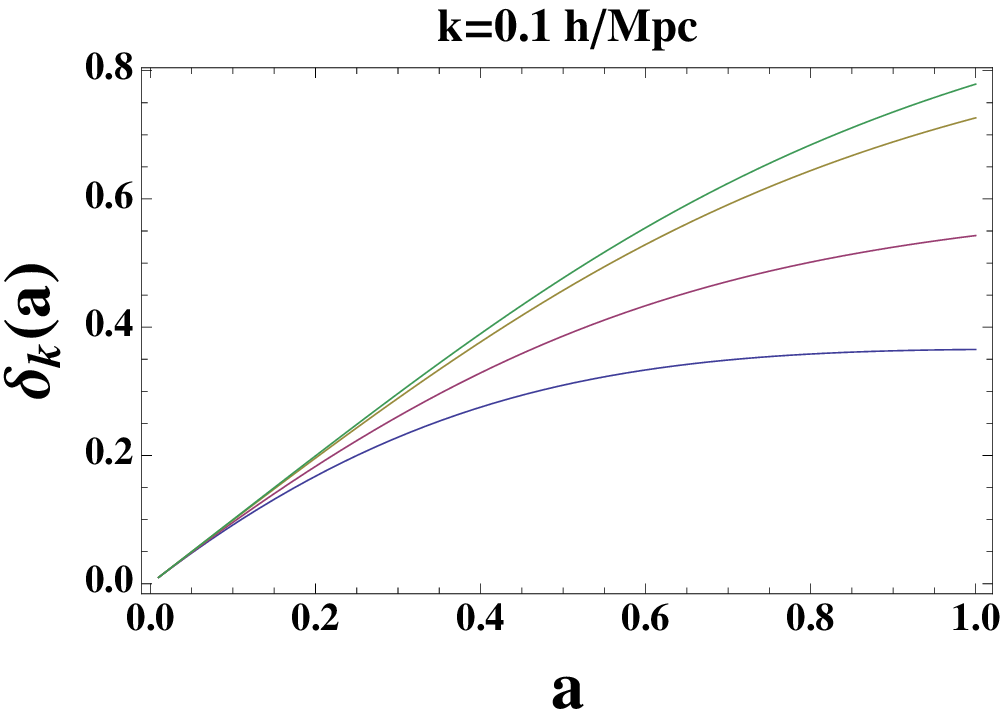}}
\end{center}
\caption{The scale-dependence of linear gravitational clustering is illustrated.
The linear density contrast, $\delta$, is shown as a function of the expansion factor, $a$, 
for the non-canonical scalar field with $\alpha = 5\times 10^4, ~10^5, ~5\times 10^5$ and also for
 $\Lambda$CDM (bottom to top with $\Lambda$CDM at the
top). Two scales are considered: $k = 0.01 h/Mpc$ (left) and $k = 0.1 h/Mpc$ (right).} 
\label{fig:3}
\end{figure*}

The evolution equation for the linear density contrast of the X-fluid in (\ref{eq:mixture1}), 
namely $\delta = \frac{\rho_X - {\bar \rho_X}}{{\bar \rho_X}}$, evaluated 
on sub-horizon scales ($|{\bf k}| \gg H/c$), is given by \cite{sandvik04}

\ber
\delta_k^{''} &=& - [2+ A - 3(2w-c_{s}^2)]\delta_k^{'} \nonumber\\
&+& \frac{3}{2}\Omega_{X}(1- 6 c_{s}^2 + 8w - 3w^2)
\delta_k -{\left(\frac{k c_{s}}{aH}\right)^2}\delta_k
\eer

\noindent
where $w$, the equation of state of the X-fluid, is described by (\ref{eq:state}),
$\Omega_{X} = 8\pi G\rho_X/3H^2$,
 $A = \frac{(H^2)^{'}}{2H^2}$ and $^{'} \equiv \frac{d}{d\log a}$. We evolve this equation from the 
decoupling epoch ($a = 10^{-3}$) when it is reasonable to assume $\delta_k \sim a$ and 
$\frac{d\delta_k}{da} \sim 1$. 
 Our results are shown in figure \ref{fig:3}
 for two different scales, $k = 0.01 h/Mpc, ~0.1 h/Mpc$. 
We find that gravitational clustering in this model is {\em scale-dependent}.
 On very large scales $k \leq 0.01 h/Mpc$, scalar field models with
large values of $\alpha \geq 10^4$ 
display clustering 
identical to $\Lambda$CDM. However on smaller scales $k \geq 0.1 h/Mpc$, the density
contrast in our model is {\em suppressed} relative to $\Lambda$CDM even 
for $\alpha$ values as large as $10^5$, for which the background expansion is 
indistinguishable from $\Lambda$CDM, as demonstrated in figure 1.

We have thus demonstrated that our model is capable of mimicking the behaviour of a dark matter + vacuum energy model both with respect to cosmological expansion and gravitational clustering.
Another possibility provided by our model is that
 the non-canonical scalar comes as an add-on
to dark matter (instead of replacing it). 
This is the usual procedure adopted by models such as quintessence,
in which the matter part of the Lagrangian remains unchanged while dark energy
is sourced by a potential such as $V \propto \phi^{-\alpha}$. 
It is easy to see that the expansion history in such a model (consisting of conventional
 dark matter and a non-canonical scalar) is
\beq
H(z) = H_0\left\lbrack \Omega_{0m} (1+z)^3 + \Omega_{0X} (1+z)^{3(1+w)} + \Omega_\Lambda\right\rbrack^{1/2}
\label{eq:hubble2}
\eeq
where the last two terms are sourced by the scalar field and $w$ is described by (\ref{eq:state}).

The clustering properties of the non-canonical scalar
 are once more given by (\ref{eqn: sound speed model}).
The new model (\ref{eq:hubble2}) therefore describes a universe filled with
the cosmological constant and 
two kinds of dark matter: the first being the usual dark matter whereas,
depending upon the value of $w \equiv c_s^2$, the second component, $\Omega_{0X}$, can
behave like a hot, warm or cold dark matter component. This could have interesting 
cosmological consequences.
For instance, as recently demonstrated in \cite{kns15}, a model with
$\Omega_{0X} \ll \Omega_{0m}$ could help
 alleviate the tension
faced by $\Lambda$CDM in simultaneously fitting CMB and weak lensing data.
A subdominant component of dark matter, like the one discussed in this paper,
 could also seed early black hole 
formation, as discussed in \cite{sawicki13}.

\section{Discussion}
\label{sec: conclusion}

In this paper we have demonstrated that a single non-canonical scalar field can
play the dual role of describing both dark matter and dark energy.
To summarize, a non-canonical scalar field rolling along a flat potential 
has a kinetic energy which decreases rapidly with time and a potential energy
which decreases much more slowly.
For large values of the  non-canonical parameter $\alpha$ in (\ref{eqn: Lagrangian}),
the kinetic energy can play the role of dark matter while the potential energy
behaves like a cosmological constant. 
The expansion history of this model therefore mimicks
$\Lambda$CDM.

On its own this result, while surprising, is not unique.
It is well known that for a given expansion history, $a(t)$,
it is always possible to reconstruct the canonical scalar field potential $V(\phi)$ which
will reproduce the expansion history precisely \cite{recon}. Therefore, in principle,
 it is possible to obtain a potential
which reproduces the $\Lambda$CDM expansion rate 
$a(t) \propto \left (\sinh{\frac{3}{2}\sqrt{\frac{\Lambda}{3}}t}\right )^{2/3}$.
However the fact that (non-oscillating) canonical scalar fields do not cluster on sub-horizon scales,
 prevents this potential
from providing a realistic portrayal of $\Lambda$CDM; also see \cite{bertacca08}.

The big advantage of non-canonical scalars arises from the fact that, for large
values of the non-canonical parameter $\alpha$, the sound speed in (\ref{eqn: sound speed model})
drops to zero. Therefore the non-canonical scalar field 
can cluster, in contrast to canonical models in which clustering is absent.

It is necessary to point out that properties similar to those possessed by
 our model have also appeared in other
discussions of unification (of dark matter and dark energy).
For instance  in \cite{scherrer04} Scherrer proposed a non-canonical model which had
 an expansion rate
exactly like $\Lambda$CDM. Our model differes from \cite{scherrer04} in two main respects:

(i) The purely kinetic Lagrangian ${\cal L}(X)$ in \cite{scherrer04} 
possesses an extremum in $X$ about which it is
expanded in a Taylor series. Our Lagrangian, on the other hand, has a power law
kinetic term with
no extremum.

(ii) The sound velocity in \cite{scherrer04} drops off as $a^{-3}$ whereas in our model
model the sound velocity is a constant and is given by (\ref{eqn: sound speed model}).

We therefore conclude that the whereas the expansion history in our model and in
\cite{scherrer04} is identical (corresponding to $\Lambda$CDM), 
the nature of gravitational clustering in these two models is rather different.
Indeed, gravitational clustering in our model is scale-dependent,
and is sensitive to the choice of $\alpha$.

\begin{figure*}[!htb]
\begin{center} 
\resizebox{220pt}{180pt}{\includegraphics{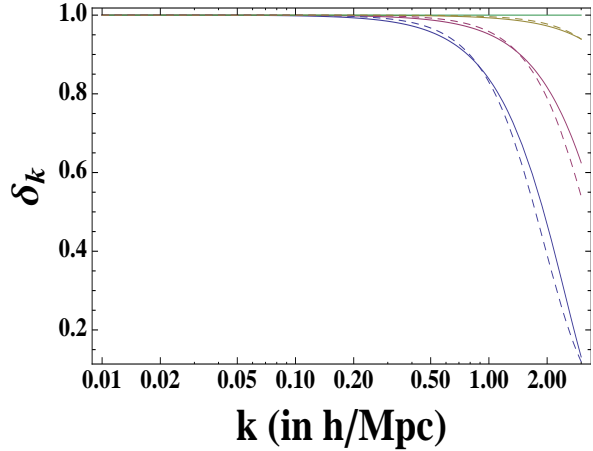}}
\end{center}
\caption{Perturbations, $\delta_k$, in the non-canonical scalar field model are shown at the present epoch.
$\delta_k$ is plotted against $k$ for $\alpha = 2\times 10^7, ~7\times 10^7, ~5\times 10^8$ 
(bottom to top,  solid lines). Perturbations in warm dark matter consisting of a sterile neutrino with
mass $= 0.5, 1, 3 ~KeV$  are also shown (bottom to top, dashed lines).
 The top most solid line corresponds to $\Lambda$CDM.} 
\label{fig:4}
\end{figure*}

In this context one should note that the value of $\alpha$ can never be infinitely large.
Consider two models characterized by $\alpha_1$ and $\alpha_2$ where 
$1 \ll \alpha_1 \ll \alpha_2$. 
Since the Jeans length in our model is
\beq
\lambda_J \sim c_s/\sqrt{G\rho}, ~~ {\rm where} ~~ c_s = \f{1}{\sqrt{2\,\alpha - 1}}~,
\label{eq:jeans}
\eeq
 it follows that the clustering properties of our
field will be sensitive to the value of $\alpha$.
Clearly gravitational clustering in the model with $\alpha_1$ 
will be inhibited on small scales relative to the model with $\alpha_2$.
This property was illustrated by figure \ref{fig:3}. It is interesting that a
similar situation arises when dark matter is sourced by an oscillating massive
canonical scalar field with mass $m$ \cite{peebles99,sahni-wang,hu00,matos00}. In this case,
as shown in \cite{hu00,zeldovich}, the Jeans length depends upon the scalar field mass as
$\lambda_J \sim (G\rho)^{-1/4} m^{-1/2}$. Ultra-light scalars are therefore able to suppress
clustering on small scales, thereby providing a resolution to the substructure and cuspy core
problems which plague standard cold dark matter.\footnote{The substructure problem
relates to the observation that CDM predicts an order of magnitude more faint galaxies
than are observed. The cuspy core problem refers to the tension between simulations of CDM,
which predict a density profile steeper than $\rho \sim 1/r$ for dark matter halos, and the much shallower
`cored' profiles  observed in individual galaxies; see \cite{sahni} and references therein.}  
One expects that a similar mechanism will operate in our model as well,
with $\alpha$ playing the role of $m$.
(A key distinction between the two models is that whereas the canonical scalar field
needs to oscillate in order to describe dark matter, the non-canonical field does not
oscillate but simply rolls along its flat potential.)

A useful analogy can also be drawn between clustering in our model and that in
particle dark matter. Consider the case when a relic particle of
mass $m$ (such as a neutralino or a sterile neutrino) plays the role of dark matter.
In this case perturbations on scales smaller than the free-streaming distance \cite{kt}
\beq
\lambda_{\rm fs} \sim 40\, {\rm Mpc} \left (\frac{m}{30 {\rm eV}}\right )^{-1}
\label{free_streaming}
\eeq
are effectively erased during the relativistic motion of the particle.

A larger value of $m$ in (\ref{free_streaming}) leads to a smaller value of 
$\lambda_{\rm fs}$.  Comparing (\ref{free_streaming}) with (\ref{eq:jeans})
we find that the role of {\em mass}, in particle dark matter models,
is played by the parameter $\alpha$ in our model.
In other words, whereas very large
values of $\alpha$
will make clustering in our model resemble cold dark matter, smaller values
of $\alpha$ will make our model closer to {\em warm dark matter}.
We demonstrate 
the similarity of our model with the sterile neutrino model for warm dark matter 
in figure \ref{fig:4}. In this figure we show $\delta_k$ for our model, obtained by solving equation (25),
and compare it with the density contrast in the warm dark matter model, as described in \cite{neutrino}; also see \cite{neutrino1}. This figure clearly demonstrates that, for suitable values of $\alpha$,
clustering in our model can be like cold or warm dark matter, even as its expansion history
mimicks $\Lambda$CDM.
The cosmological properties of our model will be examined in greater detail in a companion paper
\footnote{While we have drawn attention to the close similarity between k-essence models
and cold/warm dark matter, this analogy has been drawn at the linearized level.
At the nonlinear level these two approaches may give rise to distinct scenario's of
structure formation, as noted in \cite{sawicki13}. This issue deserves more scrutiny.}

Finally, we would like to draw attention to the following interesting
property of the non-canonical scalar which follows from (\ref{eq:phidot}).
If the value of $\alpha$ is large, then the velocity of the scalar field
freezes to an almost constant value, \ie for
 $\alpha \gg 1$, $X \to {\rm constant}$. This is completely unlike 
the behaviour of a canonical scalar ($\alpha = 1$) for which 
${\dot\phi} \propto a^{-3}$ if $V = {\rm constant}$. In the canonical
 case the universe
inflates once $V > X$, at which point since 
$a \propto e^{\int \sqrt{V} dt}$ and ${\dot\phi} \propto a^{-3}$,
the value of ${\dot\phi}$ rapidly drops to zero and
the motion of the scalar field comes to an abrupt halt. 
Not so for the non-canonical scalar
which can continue to roll along a flat potential even as
the universe inflates.

This may have interesting consequences. 
It has been postulated  \cite{banks85,review}
that the vacuum energy is not unchanging but
a dynamical quantity whose value changes abruptly during the course of a phase transition.
Such behaviour is mimicked by the step-like potential (\ref{eq:step}).
Since this potential is piece-wise flat, it can, in principle, describe a universe
which inflates twice -- once at early times, and then again at the current 
cosmological epoch.
But for this to occur, the scalar field must continue to roll along
$V$ for a protracted time interval (corresponding to the age of the universe)
 and not stop in between (as in the canonical case).
In the non-canonical context, for large values
of $\alpha$, the kinetic term 
is almost a constant, which should allow the scalar field to roll along its 
potential as the latter cascades to lower values. Such a scenario may
provide us with a model of Quintessential-Inflation and 
this possibility will be examined in greater detail in a companion paper.

\section*{Acknowledgments}

We acknowledge useful correspondence with Sanil Unnikrishnan and Shruti Thakur.




\begin{thebibliography}{99}

\bibitem{review}
V. Sahni and A.A. Starobinsky, Int. J. Mod. Phys. {\bf D9} 373 (2000);
P.~J.~E. Peebles and B. Ratra, Rev. Mod. Phys. {\bf 75} 559 (2003);
T. Padmanabhan, Phys. Rep. {\bf 380} 235 (2003);
V. Sahni, astro-ph/0202076, astro-ph/0502032;
E. J. Copeland, M. Sami and S. Tsujikawa, Int. J. Mod. Phys. {\bf D15} 1753 (2006);
R. Bousso, Gen. Relativ. Gravit. {\bf 40} (2008) 607.

\bibitem{ss06}
V. Sahni and A.A. Starobinsky, Int. J. Mod. Phys. {\bf D15} 2105 (2006).

\bibitem{bao_2014}
T. Delubac \etal, 2014, arXiv:1404.1801

\bibitem{sss14}
V. Sahni, A. Shafieloo and A.A. Starobinsky, Astrophys.J. {\bf 793} L40 (2014).

\bibitem{sarkar15}
{\em Marginal evidence for cosmic acceleration from Type Ia supernovae},
J. T. Nielsen, A. Guffanti, S. Sarkar,
 arXiv:1506.01354.

\bibitem{vms15}
{\em Cosmological Hints of Modified Gravity ?},
E. Di Valentino, A. Melchiorri, J. Silk, arXiv:1509.07501.

\bibitem{planck_2015}
{\em Planck 2015 results. XIV. Dark energy and modified gravity},
P. Ade  \etal, arXiv:1502.01590.

\bibitem{kt}
E.W. Kolb and M.J. Turner, {\em The early Universe}, Addison-Wesley Publishing Company, 1990.

\bibitem{dark_matter}
{\em Particle Candidates for Dark Matter}, John Ellis,
Invited talk presented at the Nobel Symposium, Haga Slott,
Sweden,
  August 1998, astro-ph/9812211;
{\em Non-Baryonic Dark Matter -- A Theoretical Perspective},
Leszek Roszkowski, Invited review talk at COSMO-98, the Second International Workshop on
   Particle Physics and the Early Universe, Asilomar, USA, November 15-20, 1998,
hep-ph/9903467;
A. Del Popolo, International Journal of Modern Physics D Vol. 23, No. 3 (2014) 1430005 [arXiv:1305.0456];
{\em Axion Cosmology},  David J. E. Marsh, arXiv:1510.07633.

\bibitem{sahni}
V. Sahni, Lect.Notes Phys. 653 141-180 (2004) [astro-ph/0403324]

\bibitem{mond}
{\em Modified Newtonian Dynamics as an Alternative to Dark Matter},
Robert H. Sanders, Ann.Rev.Astron.Astrophys. 40 (2002) 263-317 [astro-ph/0204521];

\bibitem{peebles99}
P.~J.~E. Peebles and A. Vilenkin, \prd {\bf 60}, 103506 (1999);
P.~J.~E. Peebles, astro-ph/0002495;

\bibitem{sahni-wang}
V. Sahni and L. Wang, \prd {\bf 62}, 103517 (2000);

\bibitem{hu00}
W.Hu, R. Barkana and A. Gruzinov, \prl {\bf 85}, 1158 (2000);

\bibitem{matos00}
T. Matos, F.S. Guzman and D. Nunes, \prd {\bf 62}, 061301 (2000);
T. Matos and L. Aryro Urena-Lopez, \prd {\bf 63}, 063506 (2001);
{\em A brief Review of the Scalar Field Dark Matter model},
Juan Maga\~na, Tonatiuh Matos, Victor Robles, Abril Su\'arez,
arXiv:1201.6107;


\bibitem{zeldovich}
M. Yu. Khlopov, B.A. Malomed and Ya.B. Zeldovich, \mnras {\bf 215}, 575 (1985).

\bibitem{CG}
A.Y. Kamenshchik, U. Moschella and V. Pasquier,
\plb {\bf 511}, 265 (2001);
N. Bilic, G.B. Tupper and R.D. Viollier \plb {\bf 535}, 17 (2002)
M. C. Bento, O. Bertolami and A. A. Sen, Phys. Rev. D, {\bf 66}, 043507 (2002).

\bibitem{tirth}
T. Padmanabhan and T. Roy Choudhury, \prd {\bf 66}, 081301 (2002).

\bibitem{CG_pert}
D. Curturan and F. Finelli, \prd {\bf 68}, 103501 (2003);
L. Amendola, F. Finelli, C. Burigana and D. Curturan,
JCAP 07(2003)005;
R. Bean and O. Dore, \prd {\bf 68}, 023515 (2003).

\bibitem{sandvik04}
H.B. Sandvik, M. Tegmark, M. Zaldarriaga and I. Waga,
\prd {\bf 69}, 123524 (2004).

\bibitem{bertacca08}
D. Bertacca, N. Bartolo and S. Matarrese,
Advances in Astronomy, Volume 2010 (2010), Article ID 904379, [arXiv:1008.0614]

\bibitem{bertacca11}
D. Bertacca, M. Bruni, O.F. Piatella and D. Pietrobon,
JCAP02(2011)018 [arXiv:1011.6669]

\bibitem{asen14}
S. Kumar and A.A. Sen, JCAP10(2014)036.

\bibitem{santiago}J.~De-Santiago, J.~L.~Cervantes-Cota and D.~Wands, Phys.~Rev.~D {\bf 87}, 023502 (2013).
\bibitem{Mukhanov-2006}
V.~Mukhanov and A.~Vikman, JCAP \textbf{0602}, 004 (2006).

\bibitem{scherrer04}
R.J. Scherrer, \prl {\bf 93}, 011301 (2004)

\bibitem{frieman}
J.A. Frieman, C.T. Hill, A. Stebbins and I. Waga,
\prl {\bf 75}, 2077 (1995).

\bibitem{scherrer2008}
S. Dutta and R.J. Scherrer,  Phys.~Rev.~D~\textbf{78}, 123525 (2008);
U. Alam, V. Sahni and A.A. Starobinsky, JCAP 0304 (2003) 002
[astro-ph/0302302].

\bibitem{tejedor_feinstein}
A. Diez-Tejedor and A. Feinstein, \prd {\bf 74}, 023530 (2006).

\bibitem{bertacca07}
D. Bertacca, S. Matarrese and M. Pietroni, Mod.Phys.Lett. {\bf A22} 2893 (2007) 
[arXiv:astro-ph/0703259]

\bibitem{sanil-2008}
S.~Unnikrishnan, Phys.~Rev.~D~\textbf{78}, 063007 (2008).

\bibitem{sanil13}
S.~Unnikrishnan, V. Sahni and  A. Toporensky, JCAP 1208 (2012) 018;
S.~Unnikrishnan and V. Sahni
JCAP 1310 (2013) 063.

\bibitem{wands12}
D. Wands, J. De-Santiago and Y. Wang, Class.Quant.Grav. 29, 145017 (2012)
[arXiv:1203.6776 [astro-ph.CO]].

\bibitem{Bardeen-1980}
J.~M.~Bardeen, Phys.\ Rev.\ D {\bf 22}, 1882 (1980).

\bibitem{Kodama-1984}
H.~Kodama and M.~Sasaki, Prog.\ Theor.\ Phys.\ Suppl.\ {\bf 78}, 1 (1984).

\bibitem{Mukhanov-1992}
V.~F.~Mukhanov, H.~A.~Feldman and R.~H.~Brandenberger, Phys.\ Rep.\ {\bf 215}, 203 (1992).

\bibitem{Garriga-1999}
J.~Garriga and V.~F.~Mukhanov, Phys.\ Lett.\ B\ {\bf 458}, 219 (1999)

\bibitem{kns15}
M. Kunz, S. Nesseris and I. Sawicki, arXiv:1507.01486

\bibitem{sawicki13}
I. Sawicki, V. Marra and W. Valkenburg, Phys. Rev. D, {\bf 88}, 083520 (2013),
[arXiv:1307.6150]

\bibitem{recon}
J.D. Barrow, \plb {\bf 235}, 40 (1990);
A.A. Starobinsky, JETP Lett. {\bf 68}, 757 (1998).

\bibitem{neutrino}
K. Abazajian Phys. Rev. D, {\bf 73}, 063513 (2006) [astro-ph/0512631]

\bibitem{neutrino1}
R.M. Dunstan, K. Abazajian, E. Polisensky and M. Ricotti, arXiv:1109.6291;
C. Destri, H.J. de Vega and N.G. Sanchez, \prd {\bf 88}. 083512 (2013);
M. Lattanzi, R. A. Lineros and M. Taoso, New J. Phys. {\bf 16} 125012 (2014)
[arXiv:1406.0004];
U. Maio and M. Viel, MNRAS 446, 2760–2775 (2015) [arXiv:1409.6718].

\bibitem{banks85}
T. Banks, Nucl. Phys. {\bf 249} (1985) 332.


\end{thebibliography}
\end{document}